\documentclass[referee]{cjaa}           % referee version: for submission

\usepackage{graphicx}                   %for PS/EPS graphics inclusion, new
\input{epsf.sty}                        %for PS/EPS graphics inclusion, old

\begin{document}

   \title{Masses of Black Holes in the Universe
%\footnotetext{$*$ Supported by the National Natural Science Foundation of China.}
}
%   \subtitle{I. Place Your Subtitle Here}

%% \volnopage is not defined at CAMK system.  Define it as an empty
%% macro.
\ifx \volnopage \undefinedmacro
    \def \volnopage#1{}
\fi

   \volnopage{Vol.0 (200x) No.0, 000--000}      %%preserved for Editor. DOn't remove!
   \setcounter{page}{1}          %%starting page, preserved for Editor. DOn't remove!

   \author{Janusz Zi\'o{\l}kowski
   %   \inst{}\mailto{}
%% Please move "\mailto{}" to the corresponding author of the paper
      }
   \offprints{J. Zi\'o{\l}kowski}                   %% is disabled in fact

   \institute{Copernicus Astronomical Center, ul. Bartycka 18, 00-716 Warsaw, Poland\\
             \email{jz@camk.edu.pl}
%% offprint requests will be sent. Note to pair \mailto{} with \email{}
          }

   \date{Received~~2007 month day; accepted~~2007~~month day}

   \abstract{
The different methods of determination of black holes (BHs) masses
are presented for three classes of BHs observed in the Universe:
stellar mass BHs, intermediate mass BHs (IMBHs) and supermassive
BHs (SBHs). The results of these determinations are briefly
reviewed: stellar mass BHs are found in the range of about 3 to
about 20 $M_\odot$, IMBHs in the range of a few hundreds to a few
tens of thousands $M_\odot$ (the determinations are much less
precise for these objects) and SBHs in the range of about $3
\times 10^5 M_\odot$ to about $6 \times 10^{10} M_\odot$.
   }

   \authorrunning{J. Zi\'o{\l}kowski}            %author_head in even pages
   \titlerunning{Masses of Black Holes in the Universe}  % title_head in odd pages

   \maketitle
%% The author head (on even pages) and the title head (on odd pages) will be
%% automatically extracted from \author and \title. Whenever the title is too long,
%% you will be asked to supply a shorter one by inserting either \authorrunning{} or
%% \titlerunning{} before \maketitle. Anyway, you can specify your own heads.
%
%________________________________________________ sections below
%
\section{Introduction}           %% first-level sections will be auto-capitalized
\label{sect:intro}
%\hspace{15pt}%                   %% preserved for Editor

Black holes (BHs) observed in the Universe can be classified into
three groups: stellar mass BHs, intermediate mass BHs (IMBHs) and
supermassive BHs (SBHs). The case for the existence of IMBHs is
still substantially weaker than the case for the two other groups
(the existence of which is, practically, beyond any dispute).
Therefore, the mass estimates for these two groups are
substantially more precise than for IMBHs. This paper is devoted
to a brief review of the different techniques used to estimate the
masses of BHs and to the brief presentation of the results.

%% ChJAA editors DID NOT use \cite{} for citation, \ref and \label for
%% cross-references of Table/Figure in publication version.
%% ChJAA editors prefered you giving a citation as 'Michel et al. 1992', and
%% writting Table~1 or Fig.~1 and so forth. However, that will make authors
%% inconvenient in adjusting/adding/removing text, tables or figures. Anyway,
%% authors can use \cite, \citep and \citet as widely used in other journals.
%% ChJAA editors are moving to use a more flexible LaTeX source.

\section{Stellar Mass Black Holes}
\label{sect:smbh}
%\hspace{15pt}%                   %% preserved for Editor

As of today, the stellar mass black holes, essentially, are
observed only in the X-ray binaries (XRBs). Only a few rough
estimates are available now for single black hole candidates from
microlensing events (see the next subsection).

\subsection{Black Holes in X-Ray Binaries}
%\hspace{15pt}%                   %% preserved for Editor

At present, 58 black hole candidates are known among compact
components of XRBs. The masses are determined for 24 of them (see
Table 1). Some of these determinations belong to the most precise
mass estimates ever derived for any black hole. Below, I briefly
summarize the technics used to obtain these estimates.

$\bullet$ {\bf The mass function}

This is the most important observational parameter used to
constrain the mass of the compact component. The mass function
$f(M_x)$ is calculated from the radial velocities of the optical
companion:

$$f(M_x) = 1.0385 \times 10^{-7} K_{opt}^3P  \,M_\odot \eqno(1)$$

\noindent where $K _{opt}$ is the semiamplitude of the radial
velocities of the absorption lines of the optical component (in
km/sec) and P is the orbital period (in days). The mass function
is related to the masses of both components by:

$$f(M_x) = M_x^3 sin^3i/(M_{opt} + M_x)^2 \eqno(2)$$

\noindent where $M_x$ and $M_{opt}$ are the masses of the compact
and the optical components and $i$ is the inclination of the
orbit. The value of $f(M_x)$ gives an absolute lower limit to the
compact component mass. To obtain a more precise value (not just
the lower limit), we have to estimate the mass ratio $q =
M_{opt}/M_x$ and the inclination of the orbit $i$.

$\bullet$ {\bf The rotational broadening of the absorption lines
of the optical component}

From the measurements of the lines, we determine the projected
rotational velocity at  the equator of the optical component
$v_{rot} sin\, i$. Assuming the corotation of the optical
component with the orbital motion (for a Roche lobe filling
component, it is a very good assumption), we have:

$$ v_{rot} sin\, i = 0.46 K_{opt} (q/(1+q))^{1/3}  \eqno(3)$$

\noindent With the help of this equation, one can determine the
mass ratio $q$. The intrinsic width of the absorption lines is
very small ( $\sim$ few km/sec), while the typical rotational
broadening is of the order of few tens km/sec and can be,
relatively easily, measured. This method has been applied, with a
substantial success, to many BH systems (see Orosz, 2003).

$\bullet$ {\bf  The radial velocities of the emission lines of the
accretion disc}

The shifts of these lines reflect the orbital motion of the
compact component and so permit us to determine the mass ratio
directly:

$$ q = K_{em}/ K_{opt}   \eqno(4)$$

\noindent where $K _{em}$ is the semiamplitude of the radial
velocities of the emission lines. This method of determining the
mass ratio is completely independent of the previous one.
Unfortunately, it is rather uneasy to implement, since the
emission lines are very broad ($\sim$ 2000 km/sec), while the
orbital shifts are of the order of few tens km/sec.

$\bullet$ {\bf  The amplitude of the ellipsoidal light variations}

Due to filling of the Roche lobe, the optical component is tidally
distorted and due to rotation it exhibits the ellipsoidal light
variations (double sinusoid per orbital period if the rotation is
synchronous). The amplitudes of these variations in V and I are
given by:

$$ \Delta V = 0.26 \,sin^2 i /(1+q)  \eqno(5)$$

$$ \Delta I = 0.24 \,sin^2 i /(1+q)  \eqno(6)$$

\noindent As may be seen, the dependence on the mass ratio $q$ is
rather weak (for most of the BH XRBs $q$ falls in the range 0.05
$\div$ 0.2) and, therefore, the ellipsoidal light variations
provide us with a valuable information about the inclination $i$
(even, if we do not know $q$). In practice, the procedure is not
that simple, since the optical light is usually (even in the
quiescence) contaminated by the residual contribution from the
accretion disc. To make the situation worse, this residual
contribution is frequently variable. Extraction of the true value
of the inclination requires, often, a very careful modeling
(Froning and Robinson 2001, Gelino et al. 2001).

$\bullet$ {\bf  The mass-spectral type relation for the optical
component}

The typical optical components of BH XRBs are the lower main
sequence stars, which satisfy reasonably well the mass-spectral
type relation. This relation may be used to estimate the mass of
the compact component, if the mass ratio is unknown. Since the
mass of the optical component is usually quite small (below 1
$M_\odot$), even substantial uncertainty (say factor of two) does
not influence dramatically the mass estimate for the compact
component.

$\bullet$ {\bf High frequency QPOs and X-ray spectra}

Using this method, one can estimate simultaneously the spins and
the masses of black holes. The spin can be estimated from careful
analysis of either continuum X-ray spectrum (Shafee et al., 2006;
Davis et al., 2006; McClintock et al., 2006) or spectral X-ray
lines (Miller et al., 2002; Miller et al., 2004; Miller et al.,
2005; Miniutti et al., 2004). The first approach gives, at
present, the more reliable results. Once the spin is known, we can
use the observed high frequency QPOs to derive the mass of a given
black hole. To do so, one has to apply one of the theories of high
frequency QPOs in the systems containing accreting black holes.
The most successful one seems to be, at the moment, the parametric
epicyclic resonance theory (Abramowicz \& Klu\'zniak 2001,
Abramowicz et al. 2004, Klu\'zniak et al. 2004, Lee et al. 2004,
T\"{o}r\"{o}k et al. 2005). For a brief summary of this theory,
the reader is referred to e.g. Zi\'o{\l}kowski (2007).

%\vspace{8mm}

The results of mass estimates carried out with the help of all the
technics described above are given in Table 1.

\begin{table}[h!]
\begin{center}
\centerline{\bf Tab. 1 $-$ Masses of Black Holes in X-Ray
Binaries} \nobreak \vspace{8mm}
%\nopagebreak

%\moveleft 12mm
%\vbox{
\begin{tabular}{|rcl|l|l|r|r|c|r|}
\hline &&&&&&&&\\
\multicolumn{3}{|c|}{Name}&\multicolumn{1}{|c|}{$P_{\rm
orb}$}&\multicolumn{1}{|c|}{Opt. Sp} &\multicolumn{1}{
|c|}{X$-$R}&\multicolumn{1}{|c|}{C}&\multicolumn{1}{|c|}{$M_{\rm
BH}$/ M$_\odot$}&\multicolumn{1}{|c|}{Ref}\\ &&&&&&&&\\ \hline
&&&&&&&&\\ Cyg X\hspace*{-2.4ex}&$-$&\hspace*{-2.4ex}1&5$^d$6&O9.7
Iab&pers&$\mu$Q&20 $\pm$ 5&1\\ LMC
X\hspace*{-2.4ex}&$-$&\hspace*{-2.4ex}3&1$^d$70&B3 V&pers&&6
$\div$ 9&\\ LMC
X\hspace*{-2.4ex}&$-$&\hspace*{-2.4ex}1&4$^d$22&O7$-$9 III&pers&&4
$\div$ 10&\\
SS\hspace*{-2.4ex}&&\hspace*{-2.4ex}433&13$^d$1&$\sim$ A7
Ib&pers&$\mu$Q&4.4$\pm$ 0.8&2\\
LS\hspace*{-2.4ex}&&\hspace*{-2.4ex}5039&3$^d$906&O7f
V&pers&$\mu$Q&2.7$\div$ 5.0&3\\ XTE
J1819\hspace*{-2.4ex}&$-$&\hspace*{-2.4ex}254&2$^d$817&B9
III&T&$\mu$Q&$6.8 \div 7.4$&\\ GX
339\hspace*{-2.4ex}&$-$&\hspace*{-2.4ex}4&1$^d$756&F8$-$G2
III&RT&$\mu$Q&$\ga$ 6&4\\ GRO
J0422\hspace*{-2.4ex}&+&\hspace*{-2.4ex}32&5$^h$09&M2 V&T&&4 $\pm$
1&\\ A 0620\hspace*{-2.4ex}&$-$&\hspace*{-2.4ex}00&7$^h$75&K4
V&RT&&11 $\pm$ 2&\\ 2S
0921\hspace*{-2.4ex}&$-$&\hspace*{-2.4ex}630&9$^d$01&K0
III&pers&&1.7 $\div$ 4.3&5,6\\ GRS
1009\hspace*{-2.4ex}&$-$&\hspace*{-2.4ex}45&6$^h$96&K8 V&T&&4.4
$\div$ 4.7&\\ XTE
J1118\hspace*{-2.4ex}&+&\hspace*{-2.4ex}480&4$^h$1&K7$-$M0
V&T&&$8.5 \pm$ 0.6&7\\ GS
1124\hspace*{-2.4ex}&$-$&\hspace*{-2.4ex}684&10$^h$4&K0$-$5
V&T&&7.0 $\pm$ 0.6&\\ GS
1354\hspace*{-2.4ex}&$-$&\hspace*{-2.4ex}645&2$^d$54&G0$-$5
III&T&&$>$ 7.4 $\pm$ 0.5&8\\ 4U
1543\hspace*{-2.4ex}&$-$&\hspace*{-2.4ex}475&1$^d$12&A2 V&RT&&8.5
$\div$ 10.4&\\ XTE
J1550\hspace*{-2.4ex}&$-$&\hspace*{-2.4ex}564&1$^d$55&K3
III&RT&$\mu$Q&10.5 $\pm$ 1.0&\\ XTE
J1650\hspace*{-2.4ex}&$-$&\hspace*{-2.4ex}500&7$^h$63&K4
V&T&$\mu$Q &4.0 $\div$ 7.3&9\\ GRO
J1655\hspace*{-2.4ex}&$-$&\hspace*{-2.4ex}40&2$^d$62&F3$-$6
IV&RT&$\mu$Q&6.3 $\pm$ 0.5&\\ H
1705\hspace*{-2.4ex}&$-$&\hspace*{-2.4ex}250&12$^h$54&K5 V&T&&5.7
$\div$ 7.9&\\ GRO
J1719\hspace*{-2.4ex}&$-$&\hspace*{-2.4ex}24&14$^h$7&M0$-$5
V&T&&$>$ 4.9&10\\ XTE
J1859\hspace*{-2.4ex}&+&\hspace*{-2.4ex}226&9$^h$16&$\sim$ G
5&T&&$8 \div 10$&\\ GRS
1915\hspace*{-2.4ex}&+&\hspace*{-2.4ex}105&33$^d$5&K$-$M
III&RT&$\mu$Q&14 $\pm$ 4&\\ GS
2000\hspace*{-2.4ex}&+&\hspace*{-2.4ex}251&8$^h$3&K5 V&T&&7.1
$\div$ 7.8&\\ GS
2023\hspace*{-2.4ex}&+&\hspace*{-2.4ex}338&6$^d$46&K0 IV&RT&&10.0
$\div$ 13.4&\\ &&&&&&&&\\ \hline
\end{tabular}\end{center}

\end{table}

\vspace{13mm}

%\nopagebreak

{\small NOTES:\vspace{2mm}\\
%\vspace{4mm}
$P_{orb}$ $-$ orbital period \hspace*{27ex} Ref $-$ references\\
Opt. Sp $-$ optical spectrum \hspace*{20ex} T $-$ transient\\ X-R
$-$ X-ray variability \hspace*{24.2ex} RT $-$ recurrent
transient\\ C $-$ comments \hspace*{33.8ex} pers $-$ persistent\\
$M_{\rm BH} -$ mass of black hole component \hspace*{10.3ex}
$\mu$Q $-$ microquasar\\ The errors or ranges for $M_{\rm BH}$ are
in most cases quoted after original references. The detailed
discussion of these estimates is given in Zi\'o{\l}kowski
(2003).\\}

{\small REFERENCES:\vspace{2mm}\\ Most of the references are given
in Zi\'o{\l}kowski (2003). Additional references are: (1)
Zi\'o{\l}kowski (2005), (2) Hillwig and Gies (2006), (3) Casares
et al. (2006), (4) Hynes et al. (2003), (5) Shahbaz et al. (2004),
(6) Jonker et al. (2005), (7) Gelino et al. (2006), (8) Casares et
al. (2004), (9) Orosz et al. (2004), (10) Masetti et al. (1996).}

\vspace{10mm}

\subsection{Black Hole Candidates from Microlensing Events}
%\hspace{15pt}%                   %% preserved for Editor

Among several hundreds of microlensing events observed so far
there are about 30 so called "paralax events". These events are
long enough to show the magnification fluctuations, reflecting the
orbital motion of the Earth around the Sun. This effect permits to
calculate the "microlensing parallax" which is a measure of the
relative transverse motion of the lens with respect to the
observer. Assuming standard model of the Galactic velocity
distribution, one is then able to perform a likelihood analysis,
which permits to estimate the distance and the mass of the lens.
With the help of the above analysis, three long events were
selected as, possibly, caused by black hole lenses. The candidates
are: MACHO-98-BLG-6 (probable mass of the lens $\sim 3 \div 13
M_\odot$, Bennett et al., 2002a), MACHO-96-BLG-5 (probable mass of
the lens $\sim 3 \div 16 M_\odot$, Bennett et al., 2002a) and
MACHO-99-BLG-22 = OGLE-1999-BUL-32 (probable mass of the lens
$\sim 100 M_\odot$, Bennett et al., 2002b). Only the last of them
seems to be a robust candidate. I will also add, that Paczy\'nski
(2003) promised more BH lenses from OGLE III project in some 2
$\div$ 3 years. OGLE III detects currently more than 500 events
per year and, among them, some 20 $\div$ 30 paralax events. Based
on the present (rather poor) statistics, we might expect that few
of them (per year) should be BHCs. However, as no new firm
detections were reported so far, it seems, that Paczy\'nski's
prediction was slightly overoptimistic.

\section{Intermediate Mass Black Holes}
\label{sect:imbh}
%\hspace{15pt}%                   %% preserved for Editor

The case for the existence of intermediate mass black holes
(IMBHs) is still not very strong but slowly it gets stronger.
There are two classes of objects where we expect IMBHs to be
present: some ultraluminous compact X-ray sources (ULXs) and
centers of some globular clusters (GCs).

\subsection{Ultraluminous X-Ray Sources (ULXs)}

The term "ULXs" is probably a sort of an umbrella covering several
different classes of objects. However, the evidence is growing
that one of these classes is, most likely, a class of XRBs
containing IMBHs.

Below, I briefly list the arguments supporting this hypothesis:

$\bullet$ Some of them {\bf are confirmed XRBs}

$\bullet$ Some of these XRBs {\bf must be massive XRBs} since they
contain young massive optical components (O8 V in M 81 X-1 (Liu et
al., 2002), mid-B sg in  M 101 X-1 (Kuntz, 2005), B0 Ib in NGC
5204 X-1 (Liu et al., 2004))

$\bullet$ Many of the ULXs are found in star forming regions and
young stellar clusters.

$\bullet$ Many of them exhibits X-ray variability on time scales
hours to years.

$\bullet$ X-ray emission of some ULXs shows QPOs on time scales of
few to few tens seconds, consistent with accretion discs around
IMBHs.

$\bullet$ X-ray spectra of many ULXs are consistent with
relatively cool accretion discs around IMBHs.

$\bullet$ Energy input into giant ionization nebulae surrounding
many ULXs exclude substantial beaming (Ho II X-1, M 81 X-9, M 81
X-6, NGC 1313 X-2, NGC 1313 X-1, NGC 5408 X-1, IC 342 X-1, NGC
5204 X-1 and others). For a discussion of these objects see Pakull
and Mirioni (2003) and Miller et al. (2005).

\subsubsection{Mass estimates for the compact components of ULX
XRBs}

The mass of such objects can be estimated from the X-ray
luminosities and the QPO frequencies (if observed). No estimate
based on the mass function is available so far.

\subsubsection{M82 X-1 -- the strongest ULX candidate for containing
an IMBH}

The observed X-ray emission of this source corresponds to an
isotropic luminosity of $(2.4 \div 16) \times 10 ^{40}$ ergs/sec
(which corresponds to the mass of $150 \div 1000 M_\odot$, if the
source emits at the Eddington level). The source exhibits 0.054 Hz
and 0.114 Hz QPOs. The source is, most likely, an X-ray binary
with an orbital period of $\sim$ 62 days. It belongs to the young
stellar cluster MGG-11 (7 to 12 Myr old).

The analysis of these properties leads to the conclusion that,
most likely, the system contains an IMBH of 200 to 5000 $M_\odot$
accreting from a giant of $\sim 25 M_\odot$, filling its Roche
lobe (Patruno et al., 2006).

The mass estimates for two other probable ULX IMBHs are given in
Table 2.

\subsection{Globular Clusters}

Modeling of the gravitational field in the central regions of some
GCs indicates that they contain fairly massive ($\sim 10^3$ to
$10^4 M_\odot$) compact objects. It is likely that these objects
are IMBHs (although, in some cases, a very dense cluster of
neutron stars cannot be ruled out).

\subsubsection{Mass estimates for the compact objects in the centers of some GCs}

The principal method is the analysis of the brightness profiles of
the central regions of GCs. Additional information might be
obtained from the estimate of the velocity dispersion in the cores
of the clusters.

The useful parameter for detecting a probable presence of a
central black hole during the preliminary analysis of the
brightness profiles is the ratio of core radius to the half mass
radius $r_c/r_h$. Trenti (2007) analyzed the dynamical evolution
of a GC under a variety of initial conditions. She found, that for
a cluster consisting initially from single stars only, the final
(after relaxation) value of $r_c/r_h$ was $\sim$ 0.01, for a
cluster containing 10 \% of binaries this value was $\sim$ 0.1,
but fot the cluster containing an IMBH the value of $r_c/r_h$ was
$\sim$ 0.3. These results confirmed earlier conclusions that IMBH
clusters have expanded cores. Trenti considered subsequently 57
dynamically old (relaxed) GCs and found that for at least half of
them the value of $r_c/r_h$ is $\ga$ 0.2, which implies the
presence of an IMBH. It seems, therefore, that a substantial
fraction of old GCs contains IMBHs. The case, however, is not
closed, since a very dense cluster of neutron stars can mimic the
gravitational potential of an IMBH.

More detailed analysis of the brightness profiles leads to some
quantitative estimates of the masses of the probable central BHs.
Some of these estimates are given in Table 2.

As far, as the velocity dispersion is concerned, it was noted
(Gebhardt et al., 2002),that GCs obey the relation (or, rather, an
extension of it) between the velocity dispersion in the core and
the mass of the central BH, found earlier for the galaxies. The
relation might be useful for preliminary estimates of the central
black hole mass, if the data about velocity dispersion are
available.

Finally, one should mention that other techniques were also used
to estimate the central BHs masses in the stellar clusters. This
includes the kinematics of milisecond pulsars (NGC 6752, Ferraro
et al., 2003) and kinematics of the massive hot stars in the
central region (IRS 13, Maillard et al., 2004). The results are
also given in Table 2.

\subsection{An ULX in a Globular Cluster?}

At the end of this section, one should mention an object that
might be an ULX in the GC. This object is a bright X-ray source in
the unnamed GC which belongs to the Virgo Cluster giant elliptical
galaxy NGC 4472 (Maccarone et al., 2007). The source luminosity is
$\sim 4 \times 10 ^{39}$ ergs/sec (which corresponds to the mass
of $\sim 25 \div 30 M_\odot$, if the source emits at the Eddington
level). The source exhibits X-ray luminosity variability by a
factor of 7 in a few hours, which excludes the possibility that
the object is several neutron stars superposed. It seems likely
that the GC in question contains an ULX, which harbors a fairly
massive BH (although, perhaps, not an IMBH yet).

\begin{table}[h!]
\begin{center}
\centerline{\bf Tab. 2 $-$ {Masses of some intermediate mass black
holes}}

  \vspace{8mm}
  \begin{tabular}{|r|r|r|}
\hline &&\\
\multicolumn{1}{|c|}{Name}&\multicolumn{1}{|c|}{$M_{\rm
BH}$/M$_\odot$}&\multicolumn{1}{|c|}{Ref}\\ &&\\

\hline \multicolumn{3}{|c|}{Ultraluminous X-ray sources}\\ \hline
&&\\

M 82 X-1&200 $\div$ 5000&1\\

MCG 03-34-63 X-1&$\ga$ 2000&2\\

Cartwheel N.10&$\ga$ 1000&3\\ &&\\

\hline \multicolumn{3}{|c|}{Stellar clusters}\\ \hline &&\\

G1&$\sim$ 20 000&4\\

M15&$\sim$ 2000&5\\

$\omega$ Cen&$\sim$ 50 000&6\\

NGC 6752&few$\times (100 \div 1000)$&7\\

IRS 13&$\sim$ 1300&8\\

&&\\ \hline

  \end{tabular}\end{center}
\vspace{13mm}

{\small REFERENCES:\vspace{2mm}\\ (1) Patruno et al. (2006), (2)
Miniutti et al. (2006), (3) Wolter et al. (2006), (4) Gebhardt et
al. (2005), (5) Gerssen et al. (2003), (6) Noyola et al. (2006),
(7) Ferrano et al. (2003), (8) Maillard et al. (2004).}

\vspace{10mm}

\end{table}

\section{Supermassive Black Holes}
\label{sect:sbh}
%\hspace{15pt}%                   %% preserved for Editor

The term supermassive black holes (SBHs) is used for BHs with the
masses in the range $\sim 10^5$ to $10^{11} M_\odot$. Initially,
such objects were believed to reside only in the centers of Active
Galactic Nuclei (AGNs). During the last decade, the evidence was
accumulated, which indicates, that virtually every "normal" galaxy
contains in its center a SBH.

\subsection{Methods of Mass Estimates for Black Holes in the
Centers of Galaxies}

At present, the four methods, listed below, are being used:

$\bullet$ {\bf Kepler's law}

$\bullet$ {\bf ${\bf M_{\rm {\bf BH}} - M_{\rm {\bf bulge}}}$
relation}

$\bullet$ {\bf "Reverberation" method}

$\bullet$ {\bf "Variance" method}

\subsection{Kepler's law}

This method is based on measuring the motions of the objects
following the Keplerian orbits around the central BH. The objects
might be either individual stars or stellar aggregates or water
masers in the accretion disc. This method produces the most
precise estimates, especially, if applied to individual stars or
water masers.

In particular, the mass of Sgr A$^*$, the SBH in the center of our
Galaxy was estimated from the motions of the several stars. In
this case, both radial velocities and astrometric positions could
be measured. These observations demonstrated that the common focus
of the elliptical stellar orbits must harbor an invisible (except
the rare X-ray or IR flares) mass of (3.6 $\pm$ 0.4)$\times 10^6$
M$_\odot$ ((Ghez et al., 2003). The pericenter distance of one of
the stars (S0-16) is only about 8 light hours, so the size of the
invisible object must be smaller. It could be nothing but BH.

For our twin galaxy M31, the orbits of individual stars could not
be used and only the kinematics of stellar aggregates was
investigated. From the complex modeling of the gravitational
potential, the value 1.4$\times 10^8$ M$_\odot$ was derived for
the mass of the central BH.

For a handful of nearby galaxies water maser sources were observed
within the accretion disc (sometimes a few different sources at
different distances from the center for a given galaxy). Their
orbital motions yield quite precise estimate of the central mass.
As an example, one may quote the galaxy NGC 4258 for which the
value (3.9 $\pm$ 0.1)$\times 10^7$ M$_\odot$ was found for the
mass of the central BH (Herrnstein et al., 1999).

\subsection{${\bf M}_{\rm {\bf BH}} - {\bf M}_{\rm {\bf
bulge}}$ {\bf relation}}

Magorrian et al. (1998) found an empirical relation between the
mass of the bulge of the galaxy and the mass of the central BH.
The more recent version of this relation may be found e.g. at
H\"{a}ring and Rix (2004). As the mass of the bulge is estimated
from the stellar velocities dispersion, this relation is, in fact,
the relation between the stellar velocities dispersion and the
mass of the central BH (which is obeyed also by some GCs, as was
mentioned in the section 3.2.1). This relation may be quantified
as:

$$ M_{\rm BH} = 1.35 \times 10^8 (\sigma/200 \,km/s)^{4.02}
\,M_\odot  \eqno(7)$$

\subsection{"Reverberation" method}

This method can be applied in the case of active galaxies, where
the previous method cannot be used (the emission from the active
nucleus overshines the stellar component and the velocity
dispersion which is based on stellar absorption lines cannot be
determined). The reverberation method is essentially based on
Kepler's law ($M_{\rm BH} = v^2 R/{\rm G}$) applied, in this case,
to broad line region (BLR). The velocity $v$ of the matter in this
region is estimated from the width of the H${\beta}$ emission line
and the distance from the center $R$ from the time delay of the
variability of the emission from BLR with respect to that from the
central region. Since both the radius $R$ and the velocity $v$ (or
the H${\beta}$ width $\Delta \lambda$) were found empirically to
scale with the total luminosity of the nucleus, the relevant
formula can be simplified to (Kaspi et al., 2000):

$$ M_{\rm BH} = 5.71 \times 10^7 L_{44}^{\,\,\,\,\,\,0.545}
\,M_\odot \eqno(8)$$

\noindent where $L_{44}$ is the luminosity in the units of
$10^{44}$ erg/s.

This method was successfully applied to many active galaxies.
Unfortunately, its accuracy is rather low, since the result
depends on the inclination of the plane of BLR (which is unknown)
and on the assumption about the geometry of motions of the gas in
BLR (which probably is not exactly keplerian).

\subsection{"Variance" method}

This method (Papadakis et al., 2004; Niko\l ajuk et al., 2004) is
based on the analysis of the variability in the X-ray band (long
sequences of good quality observations are necessary). The mass of
the central object can be estimated from the high frequency break
in the power density spectrum (the frequency scales approximately
linearly with the mass. This method is completely independent of
the previous one. The new calibration of the relevant formula
(Niko\l ajuk et al., 2006) was based on the variability of the
system Cyg X-1 (and on the new determination of its mass
(Zi\'o{\l}kowski, 2005)). It is, certainly, encouraging that the
masses determined by two completely independent methods
(reverberation and variance) agree well with each other (Niko\l
ajuk et al., 2006).

Mass estimates were carried out for a large number of SBHs. Short
selection of some results is given in Table 3. It is worth
noticing that the range of masses covers more than five orders of
magnitude.

\begin{table}[h!]
\begin{center}
\centerline{\bf Tab. 3 $-$ {Masses of some supermassive black
holes}}

  \vspace{8mm}
  \begin{tabular}{|r|r|r|}
\hline &&\\
\multicolumn{1}{|c|}{Name}&\multicolumn{1}{|c|}{$M_{\rm
BH}$/M$_\odot$}&\multicolumn{1}{|c|}{Ref}\\ &&\\

\hline \multicolumn{3}{|c|}{Low mass supermassive black holes}\\
\hline &&\\

NGC 4395&$3.6 \times 10^5$&1\\

NGC 4051&$\sim 5 \times 10^5$&2\\

Sgr A$^*$&$3.6 \times 10^6$&3\\

\hline \multicolumn{3}{|c|}{High mass supermassive black holes}\\
\hline &&\\

[BH89] 1346-036&$0.9 \times 10^{10}$&4\\

LBQS 0109+0213&$1.0 \times 10^{10}$&4\\

[BH89] 0329-385&$1.3 \times 10^{10}$&4\\

2QZ J222006.7-280324&$1.4 \times 10^{10}$&4\\

[BH89] 1246-057&$1.7 \times 10^{10}$&4\\

TON 618&$6.6 \times 10^{10}$&4\\

&&\\ \hline

  \end{tabular}\end{center}
\vspace{13mm}

{\small REFERENCES:\vspace{2mm}\\ (1) Peterson et al. (2005), (2)
Shemmer et al. (2003), (3) Ghez et al., (2003), (4) Shemmer et al.
(2004).}

\vspace{10mm}

\end{table}

\begin{acknowledgements}
This work was partially supported by the State Committee for
Scientific Research grants No 4 T12E 047 27 and 1P03D 011 28.

\end{acknowledgements}

%\pagebreak


\begin{thebibliography}{99}
%% you can type \apj for ApJ, \aap for A&A, \apss for Ap&SS, etc. Please consult
%% the macro cjaa.cls. You can also find them in aasguide.tex (AASTeX for ApJ, AJ, PASP)
%% Please follow the format of ChJAA's reference list


\bibitem[2001]{abra01} Abramowicz, M.A., Klu\'zniak, W., 2001, \aap, 374, L19

\bibitem[2004]{abra04} Abramowicz, M.A., Klu\'zniak, W., McClintock, J.E., Remillard, R.A., 2004, \apj,  609, L63

\bibitem[2002a]{benn02a} Bennett D.P., Becker A.C., Quinn J.L. et al., 2002a, \apj, 579, 639

\bibitem[2002b]{benn02b} Bennett D.P., Becker A.C., Calitz J.J., Johnson B.R., Laws C., Quinn J.L.,  Rhie S.H., Sutherland, W., 2002b,
   preprint (astro-ph 0207006)

\bibitem[2005]{casa05} Casares, J., Ribo, M., Ribas, I., Paredes, J.M., Marti,
J., Herrero, A.: 2005, \mnras, 364, 899

\bibitem[2004]{casa04} Casares, J., Zurita, C., Shahbaz, T., Charles, P.A., Fender, R.P., 2004, \apj, 613, L133

\bibitem[2006]{davi06}  Davis, S.W., Done, C., Blaes, O.M., 2006,
\apj, 647, 525

\bibitem[2003]{ferr03} Ferraro, F.R., Possenti, A., Sabbi, E., Lagani, P., Rood, R.T., D'Amico, N.,
Origlia, L., 2003, \apj, 595, 179

\bibitem[2001]{fron01} Froning, C.S. and Robinson, E.L., 2001, \aj 121 2212

\bibitem[2002]{gebh02} Gebhardt, K., Rich, R.M., Ho, L.C., 2002,
\apj, 578, L41

\bibitem[2005]{gebh05} Gebhardt, K., Rich, R.M., Ho, L.C., 2005,
\apj, 634, 1093

\bibitem[2006]{geli06} Gelino, D.M., Balman, S., Kiziloglu, U., Yilmaz, A., Kalemci, E., Tomsick, J.: 2006, \apj, 642, 438

\bibitem[2001]{geli01} Gelino, D.M., Harrison, T.E. and Orosz J.A., 2001, \aj, 122, 2668

\bibitem[2003]{gers03} Gerssen, J., van der Marel, Roeland P., Gebhardt,
K., Guhathakurta, P., Peterson, R.C., Pryor, C., 2003, \aj, 125,
376

\bibitem[2003]{ghez03} Ghez, A.M., Becklin, E., Duchjne, G., Hornstein, S., Morris,
M., Salim, S., Tanner, A., 2003, Astronomische Nachrichten,
Suppl.1, 324, 527

\bibitem[2004]{hari04} H\"{a}ring, N., Rix, H.-W., 2004, \apj, 604, 89

\bibitem[1999]{herr99} Herrnstein, J.R., Moran, J.M., Greenhill, L.J., Diamond,
P.J., Inoue, M., Nakai, N., Miyoshi, M., Henkel, C., Riess, A.,
1999, \nat, 400, 539

\bibitem[2006]{hill06} Hillwig, T.C., Gies, D., 2006, Bull. AAS, 38, 954

\bibitem[2003]{hyne03} Hynes R.I., Steeghs D., Casares J., Charles P.A., O'Brien K., 2003, \apj, 583, L95

\bibitem[2005]{jonk05} Jonker, P.G., Steeghs, D., Nelemans, G., van der Klis,
M., 2005, \mnras, 356, 621

\bibitem[2000]{kasp00} Kaspi, S., Smith, P.S., Netzer, H., Maoz, D., Jannuzi,
B.T., Giveon, U., 2000, \apj, 533, 631

\bibitem[2004]{kluz04} Klu\'zniak, W., Abramowicz, M.A., Kato, S., Lee, W.H., Stergioulas, N., 2004, \apj,  603, L89

\bibitem[2005]{kunt05} Kuntz, K.D., Gruendl, R.A., Chu, Y.-H., Chen, C.-H.R., Still, M., Mukai, K., Mushotzky,
R.F., 2005, \apj, 620, L31

\bibitem[2004]{lee04} Lee, W.H., Abramowicz, M.A., Klu\'zniak, W., 2004, \apj,  609, L93

\bibitem[2002]{liu02} Liu, J.-F., Bregman, J.N., Seitzer, P.,
2002, \apj, 580, L31

\bibitem[2004]{liu04} Liu, J.-F., Bregman, J.N., Seitzer, P.,
2004, \apj, 602, 249

\bibitem[2007]{macc07} Maccarone T.J., Kundu A., Zepf, S.E., Rhode, K.L., 2007, \nat,
 445, 183

\bibitem[1998]{mago98} Magorrian, J., Tremaine, S., Richstone, D., Bender,
R., Bower, G., Dressler, A., Faber, S.M., Gebhardt, K., Green, R.,
Grillmair, C., Kormendy, J., Lauer, T., 1998, \aj, 115, 2285

\bibitem[2004]{mail04} Maillard, J. P., Paumard, T., Stolovy, S.R., Rigaut,
F., 2004, \aap, 423, 155

\bibitem[1998]{mase98} Masetti, N., Bianchini, A., Bonibaker, J., della Valle, M., Vio,
R., 1998, \aap, 314, 123

\bibitem[2006]{mccl06} McClintock, J.E., Shafee, R., Narayan, R., Remillard, R.A., Davis, S.W.,
Li, L-X.: 2006, \apj, 652, 518

\bibitem[2005]{mill05} Miller, J.M., Fabian, A.C., Nowak, M.A., Lewin, W.H.G.: 2005, {\it Procs. of 10-th Marcel Grossman Meeting},
eds. Novello, M., Peres-Bergliaffa, S., Ruffini, R., World
Scientific, Singapore, (also astro-ph 0402101).

\bibitem[2002]{mill02} Miller, J.M., Fabian, A.C., Wijnands, R., Reynolds, C.S., Ehle, M., Freyberg, M.J.,
van der Klis, M., Lewin, W.H.G., Sanchez-Fernandez, C.,
Castro-Tirado, A.J.: 2002, \apj, 570, L69.

\bibitem[2004]{mill04} Miller, J. M., Homan, J., Steeghs, D., Rupen, M., Wijnands,
R., Charles, P., 2004, Bull. AAS, 36, 945

\bibitem[2005]{mill05} Miller, N.A., Mushotzky, R.F., Neff, S.G.,
2005, \apj, 623, L109

\bibitem[2004]{mini04} Miniutti, G., Fabian, A.C., Miller, J.M., 2004, \mnras,
351, 466

\bibitem[2006]{mini06} Miniutti, G., Ponti, G., Dadina, M., Cappi, M., Malaguti, G., Fabian, A.C., Gandhi,
P., 2006, \mnras, 373, L1

\bibitem[2006]{niko06} Niko\l ajuk, M., Czerny, B., Zi\'o{\l}kowski, J., Gierliñski,
M., 2006, \mnras, 370, 1534

\bibitem[2004]{niko04} Niko\l ajuk, M., Papadakis, I.E., Czerny,
B., 2004, \mnras, 350, 26

\bibitem[2006]{noyo06} Noyola, E., Gebhardt, K., Bergmann, M.,
2006, ASP Conf. Series, 352, 269

\bibitem[2003]{oros03} Orosz, J.A., 2003, Procs. of IAU Symp. 212, p. 265 (see also
astro-ph 0209041)

\bibitem[2004]{oros04} Orosz, J.A., McClintock, J.E., Remillard, R.A., Corbel, S., 2004, \apj, 616, 376

\bibitem[2003]{pacz03} Paczy\'nski B., 2003, The Future of Small Telescopes In The New Millennium.
  Volume III - Science in the Shadows of Giants, ed. T.D. Oswalt,
  Astrophysics and Space Science Library, Volume 289, Kluwer
  Academic Publishers, Dordrecht, p.303 (see also astro-ph 0306564)

\bibitem[2003]{paku03} Pakull, M.W., Mirioni, L., 2003, Revista Mexicana de Astronomía y
Astrofísica, 15, 197

\bibitem[2004]{papa04} Papadakis, I.E., 2004, \mnras, 348, 207

\bibitem[2006]{patr06} Patruno, A., Portegies Zwart, S., Dewi, J., Hopman,
C., 2006, \mnras, 370, 6

\bibitem[2005]{pete05} Peterson, B.M., Bentz, M.C., Desroches, L.-B., Filippenko, A.V., Ho, L.C., Kaspi, S., Laor, A., Maoz, D., Moran, E.C.,
Pogge, R.W., Quillen, A.C., 2005, \apj, 632, 799

\bibitem[2006]{shaf06} Shafee, R., McClintock, J.E., Narayan, R., Davis, S.W., Li, L.-X.,
Remillard, R.A., 2006, \apj, 636, L113

\bibitem[2004]{shah04} Shahbaz, T., Casares, J., Watson, C.A., Charles, P.A., Hynes, R.I., Shih, S.C., Steeghs,
D., 2004, \apj, 616, L123

\bibitem[2004]{shem04} Shemmer, O., Netzer, H., Maiolino, R., Oliva, E., Croom, S., Corbett, E., di Fabrizio,
L., 2004, \apj, 614, 547

\bibitem[2003]{shem03} Shemmer, O., Uttley, P., Netzer, H., McHardy,
I.M., 2003, \mnras, 343, 1341

\bibitem[2006]{tren06} Trenti, M., 2006, preprint (astro-ph 0612040)

\bibitem[2005]{toro05} T\"{o}r\"{o}k, G., Abramowicz, M.A., Klu\'zniak, W., Stuchlik,
  Z., 2005, \aap, 436, 1

\bibitem[2006]{wolt06} Wolter, A., Trinchieri, G., Colpi, M.,
2006, \mnras, 373, 1627

\bibitem[2003]{ziol03} Zi\'o{\l}kowski J., 2003, Frontier Objects in Astrophysics and Particle Physics
  (Procs. of the Vulcano Workshop 2002), eds. F. Giovannelli \& G. Mannocchi, Conference Proceedings,
  Italian Physical Society, Editrice Compositori, Bologna, Italy, 85, 411 (see also astro-ph 0307307)

\bibitem[2005]{ziol05} Zi\'o{\l}kowski J., 2005, \mnras, 358, 851

\bibitem[2007]{ziol07} Zi\'o{\l}kowski J., 2007, Frontier Objects in Astrophysics and Particle Physics
  (Procs. of the Vulcano Workshop 2006), eds. F. Giovannelli \& G. Mannocchi, Conference Proceedings,
  Italian Physical Society, Editrice Compositori, Bologna, Italy, 93, 251


\end{thebibliography}
\end{document}